\documentclass[12pt]{iopart}
\usepackage{iopams}
\usepackage{graphicx}
\usepackage[top=1in, bottom=1.25in, left=1.25in, right=1.25in]{geometry}%toggle to get smaller page
\usepackage[colorlinks=true, citecolor=blue, urlcolor=blue ]{hyperref}
\expandafter\let\csname equation*\endcsname\relax
\expandafter\let\csname endequation*\endcsname\relax
\usepackage{amsmath, amssymb}
\usepackage{bm}
\usepackage{bbm}
\usepackage{tabulary}% Table with text wraping in cells
\usepackage[stretch=10]{microtype} % Allows font streching 
\usepackage{empheq}
\usepackage{color}
\usepackage{tcolorbox}
\usepackage{pxfonts,txfonts}
\usepackage{tgpagella}
\usepackage[T1]{fontenc}
\newcommand{\RR}{\mathbbm{R}}
\newcommand{\CC}{\mathbbm{C}}

\newcommand{\ran}{\rangle}
\newcommand{\lan}{\langle}

\newtheorem{theorem}{Theorem}

\newtheorem{proposition}[theorem]{Proposition}

\newtheorem{observation}[theorem]{Observation}

\begin{document} 
\title[Any qubit IO can be decomposed into four incoherent Kraus operators]{Optimal decomposition of incoherent qubit channel}

\author{Swapan Rana$^1$ and Maciej Lewenstein$^{1,2}$}

\address{$^1$ ICFO -- Institut de Ciencies Fotoniques, The Barcelona Institute of Science and Technology, 08860 Castelldefels (Barcelona), Spain}
\address{$^2$ ICREA, Lluis Companys 23, E-08010 Barcelona, Spain}
\ead{swapanqic@gmail.com}
%\date{Last updated: \today}
\begin{abstract}
We show that any incoherent qubit channel could be decomposed into four incoherent Kraus operators. The proof consists in showing existence of four incoherent Kraus operators by decomposing the corresponding Choi-Jamio\l{}kowski-Sudarshan  matrix. We mention some applications of this optimal decomposition. We also show that the Kraus rank and incoherent rank are different even for qubit channel. 
\end{abstract}
%\keywords{coherence, incoherent operation, Kraus operators}
\submitto{\jpa}
\maketitle

%============================== Section 1 ==========================
\section{Introduction}
In resource theory of coherence \cite{Streltsov+2.RMP.2017}, as introduced in \cite{Baumgratz+2.PRL.2014}, one first chooses a basis for the Hilbert space corresponding to the considered quantum system. Diagonal states are then regarded as free (called 'incoherent') states of the theory. The free operations are defined in terms of incoherent operators, which transform any free state to another free state. Precisely, a quantum operation is free, termed as incoherent operation (IO), if and only if it has a Kraus decomposition with all Kraus operators being incoherent. If additionally the transpositions of all such Kraus operators are also incoherent, the operation is called strictly incoherent (SIO) \cite{Winter+1.PRL.2016}. The SIOs, being more restrictive, neither create nor use coherence \cite{Yadin+4.PRX.2016}. 

The free operations in entanglement theory---the local operations and classical communications (LOCC) \cite{Chitambar+4.CMP.2014}, have clear physical restrictions that the (possibly spatially separated) parties are allowed to act only locally, albeit, lack a decomposition in terms of Kraus operators. In contrast, the free operations in coherence theory---that is the IOs, are defined in terms of the Kraus decomposition. However, this poses a difficulty in implementing the IOs, as (besides the non-uniqueness of Kraus decomposition) the general way of implementing a quantum operation---adding an ancillary system, evolving the combined system under a suitable unitary, and finally tracing off the ancillary system, is not applicable. Although an incoherent unitary and incoherent ancilla always yield an IO, these restrictions may not be necessary. Specifically, it is not known what is the necessary and sufficient condition for the unitary and ancilla to yield IOs. But, implementation of free operations are of immense importance as it dictates the whole structure of the theory starting from state conversion.    

One way to understand the structure of IOs is to find their minimal parametrization, which necessarily involves finding the minimum number of Kraus operators. It will also allow to numerically simulate the IOs efficiently for small dimensions. To this aim, some bounds on the number of Kraus operators for the (S)IOs have been derived in \cite{Streltsov+3.PRL.2017}. However, optimality of these bounds are not known even for the simplest case of qubit IO. It was shown that every SIO can be decomposed into four Kraus operators (and this is optimal number), and every IO into five Kraus operators. A canonical representation of the Kraus operators for any incoherent qubit channel is given by \cite{Streltsov+3.PRL.2017} the set
\begin{equation}\label{Pres:canonical.IO}
\left\{\begin{pmatrix}a_{1} & b_{1}\\
0 & 0
\end{pmatrix},\begin{pmatrix}0 & 0\\
a_{2} & b_{2}
\end{pmatrix},\begin{pmatrix}a_{3} & 0\\
0 & b_{3}
\end{pmatrix},\begin{pmatrix}0 & b_{4}\\
a_{4} & 0
\end{pmatrix},\begin{pmatrix}a_{5} & 0\\
0 & 0
\end{pmatrix}\right\},
\end{equation}
where $a_{i}$ can be chosen nonnegative, while $b_{i}\in\CC$. Moreover,
it holds (for normalization) that $\sum_{i=1}^{5}a_{i}^{2}=\sum_{j=1}^{4}|b_{j}|^{2}=1$
and $a_{1}b_{1}+a_{2}b_{2}=0$. Similarly, a canonical (and optimal) representation of the Kraus operators for a qubit SIO is given by \cite{Streltsov+3.PRL.2017} the set
\begin{equation}\label{Pres:canonical.SIO}
\left\{\begin{pmatrix}a_{1} & 0\\
0 & b_{1}
\end{pmatrix},\begin{pmatrix}0 & b_{2}\\
a_{2} & 0
\end{pmatrix},\begin{pmatrix}a_{3} & 0\\
0 & 0
\end{pmatrix},\begin{pmatrix}0 & 0\\
a_{4} & 0
\end{pmatrix}\right\},
\end{equation}
where $a_{i}\geq 0$, $b_{i}\in\CC$, and (for normalization) $\sum_{i=1}^{4}a_{i}^{2}=\sum_{j=1}^{2}|b_{j}|^{2}=1$.

In the present work, we will show that every qubit IO can be decomposed into four incoherent Kraus operators, thereby giving the optimal description of qubit IOs. This decomposition will be compared with Kraus rank and some applications will be mentioned later. Throughout this manuscript, the terms '(S)IO channel' and '(S)IO operations' will be used synonymously. We will also frequently use the fact that two sets of Kraus operators $\{K_i\}$ and $\{L_i\}$, each having exactly $m$ members (possibly after adding zero operators to the set having less members) describe the same quantum channel if and only if \cite[pp.~372]{Nielsen+Chuang.CUP.2010}  there is an  $m$-by-$m$ unitary matrix $U=(u_{ij})$ such that  
\begin{equation}
\label{Eq:Equivalence.of.Kraus.operators} L_i=\sum_{j=1}^m u_{ij}K_j,\quad i=1,2,\dotsc,m\,.
\end{equation}

In what follows, we will first re-parametrize the canonical form of IO from Eq.~\eqref{Pres:canonical.IO} to \begin{equation}
\label{Pre.Canonical.IO.Old}\Lambda=\left\{\begin{pmatrix}
r \alpha _1 & \beta _1 \\
0 & 0 
\end{pmatrix},\,\begin{pmatrix}
0 & 0 \\
\alpha _1 & -r \beta _1 
\end{pmatrix},\,\begin{pmatrix}\alpha_{2} & 0\\
0 & \beta_{2}
\end{pmatrix},\,\begin{pmatrix}0 & \beta_{3}\\
\alpha_{3} & 0
\end{pmatrix},\,\begin{pmatrix}\alpha_{4} & 0\\
0 & 0
\end{pmatrix}\right\},
\end{equation}
where $r, \alpha_i\geq 0$, $\beta_i\in\CC$, and for normalization, \begin{equation}
\label{Eq:Normalization.Channel.Old} \alpha _2^2+\alpha _3^2+\alpha _4^2+\left(r^2+1\right) \alpha _1^2=\left| \beta _2\right|^2+\left| \beta _3\right|^2+\left(r^2+1\right)\left| \beta _1\right|^2=1.
\end{equation}
With this parametrization, we make the following observation.
\begin{observation}\label{Obs:alpha.beta.nonzero}
				If any of the $\alpha_i$'s or $\beta_i$'s is zero in Eq.~\eqref{Pre.Canonical.IO.Old}, the number of Kraus operators reduces to four.
\end{observation}

\noindent \textbf{Proof of Observation~\ref{Obs:alpha.beta.nonzero}}: If $\alpha_4=0$, there is nothing to prove. If $\alpha_1\beta_1=0$, then $\Lambda$ is an SIO and hence could be reduced to the canonical form in Eq.~\eqref{Pres:canonical.SIO} with at most four Kraus operators. If $\beta_2=0$,  then the operators $\binom{\alpha_2~0}{0~~0}$ and $\binom{\alpha_4~0}{0~~0}$ are scalar multiple of each other, hence one could be made zero via Eq.~\eqref{Eq:Equivalence.of.Kraus.operators}, thereby reducing the number of Kraus operators to four. Similarly, if $\alpha_3=0$, the three operators $\binom{r\alpha_1~0}{0~~0}$, $\binom{0~\beta_3}{0~~0}$, and $\binom{\alpha_4~0}{0~~0}$ are linearly dependent, so one of them could be made zero. The remaining two cases of $\alpha_2=0$ and $\beta_3=0$ are similar, thus we consider the first one.
 
 If $\alpha_2=0$, then consider the (unnormalized) unitary $U\oplus V\oplus 1$, with
 \[U=\begin{pmatrix}
 r \alpha _1 & \alpha _4 \\
 -\alpha _4 & r \alpha _1 \\
 \end{pmatrix},\,V=\begin{pmatrix}
  r \beta _1^* & -\beta _2* \\
 \beta _2 & r \beta _1 \\
 \end{pmatrix},
 \] which 
 transforms \[\left\{\binom{r\alpha _1~ \beta _1}{0~~0},\,\binom{\alpha_4~0}{0~~0}\right\}\,\text{and}\, \left\{\binom{0~~~~~0}{\alpha _1~ -r\beta _1},\,\binom{0~~~0}{0~~\beta_2}\right\}\,\text{to}\,\left\{\binom{*~~ *}{0~~0},\,\binom{0~~*}{0~~0}\right\}\,\text{and}\, \left\{\binom{0~0}{*~ *},\,\binom{0~0}{*~0}\right\}\] respectively. But then the three operators $\binom{0~~*}{0~~0}$, $\binom{0~~0}{*~~0}$, and $\binom{0~~\beta_3}{\alpha_2~~0}$ are linearly dependent, and so one of those could be made zero.\hfill$\square$

%============================== Section 2 ==========================
\section{Main result}
\begin{theorem}\label{Thm:Opt.is.5}
The optimal number of incoherent Kraus operators for an incoherent qubit channel is four. That is, every incoherent qubit channel could be decomposed into four incoherent Kraus operators and there are some which cannot be decomposed into three. Thus an incoherent qubit channel can be canonically represented as \begin{equation}
\label{Pre.Canonical.IO.Optimal}\Lambda=\left\{\begin{pmatrix}
r \alpha _1 & \beta _1 \\
0 & 0 
\end{pmatrix},\,\begin{pmatrix}
0 & 0 \\
\alpha _1 & -r \beta _1 
\end{pmatrix},\,\begin{pmatrix}\alpha_{2} & 0\\
0 & \beta_{2}
\end{pmatrix},\,\begin{pmatrix}0 & \beta_{3}\\
\alpha_{3} & 0
\end{pmatrix}\right\},
\end{equation}
where $r, \alpha_i\geq 0$, $\beta_i\in\CC$, and for normalization, \begin{equation}
\label{Eq:Normalization.Channel} \alpha _2^2+\alpha _3^2+\left(r^2+1\right) \alpha _1^2=\left| \beta _2\right|^2+\left| \beta _3\right|^2+\left(r^2+1\right)\left| \beta _1\right|^2=1.
\end{equation}
\end{theorem}

Since the four operators $\{|i\ran\lan j|\}$, $i,j=1,2$ are linearly independent (in the vector space of 2-by-2 matrices over complex numbers), the channel $\Lambda=\{|i\ran\lan j|/\sqrt{2}\}$ is an (S)IO whose number of Kraus operators could not be reduced further. Thus to prove the optimality, it suffices to show that any qubit incoherent channel, which without loss of generality given in Eq.~\eqref{Pre.Canonical.IO.Old}, could be decomposed into four incoherent Kraus operators. 

\noindent \textbf{Proof:} In view of Observation~\ref{Obs:alpha.beta.nonzero}, we can assume each $\alpha_i,\beta_i$ to be non-zero in the channel \eqref{Pre.Canonical.IO.Old}. The Choi-Jamio\l{}kowski-Sudarshan (CJS) matrix for this channel is 
\begin{equation}
\label{Exp.Choi.M} M=\frac{1}{2}\begin{pmatrix}
r^2 \alpha _1^2+\alpha _2^2+\alpha _4^2 & r \alpha _1 \beta _1^* & 0 & \alpha _2 \beta _2^* \\
r \alpha _1 \beta _1 & \left| \beta _1\right|^2+\left| \beta _3\right|^2 & \alpha _3 \beta _3 & 0 \\
0 & \alpha _3 \beta _3^* & \alpha _1^2+\alpha _3^2 & -r \alpha _1 \beta _1^* \\
\alpha _2 \beta _2 & 0 & -r\alpha _1 \beta _1 & r^2\left| \beta _1\right|^2+\left| \beta _2\right|^2 
\end{pmatrix}=\frac{1}{2}\begin{pmatrix}
a & e & 0 & g \\
e^* & b & f & 0 \\
0 & f^* & c & -e \\
g^* & 0 & -e^* & d 
\end{pmatrix}.
\end{equation}
Assuming $M\geq 0$, we want to show that $M$ admits a decomposition of the form 
\begin{equation}
\label{Eq:Decomposition.M} 2M=\begin{pmatrix}
A &.&.&g\\.&.&.&.\\.&.&.&.\\g^*&.&.&\frac{\left|g\right|^2}{A}
\end{pmatrix}+\begin{pmatrix}
.&.&.&.\\. & B & f &.\\ .& f^*& \frac{\left|f\right|^2}{B}&.\\ .&.&.&.
\end{pmatrix}+\begin{pmatrix}
a-A & e&.&.\\e^*&b-B&.&.\\.&.&c-\frac{\left|f\right|^2}{B}& -e \\.&.&-e^* &d-\frac{\left|g\right|^2}{A}
\end{pmatrix}
\end{equation} 
for some $A,B> 0$. If, additionally, each of the two diagonal blocks of the last matrix is positive semidefinite (PSD) with rank one, then the matrix $M$ is decomposed into sum of four rank one PSD matrices which are CJS matrix of the four incoherent Kraus operators of the form $\binom{*~~0}{0~~*}$, $\binom{0~~*}{*~~0}$, $\binom{*~~*}{0~~0}$, and $\binom{0~~0}{*~~*}$ respectively. These four CJS matrices will uniquely determine the parameters $r,\alpha_i,\beta_i$ in the canonical form given in  Eq.~\eqref{Pre.Canonical.IO.Optimal}.   Therefore, to make the number of (incoherent) Kraus operators four, it suffices to show that the last matrix is PSD and it has rank two. This is the case if and only if
\begin{subequations}\label{Eq.Main.unknown.A.B}
	\begin{align}
	(a-A)(b-B)&=\left(c-\frac{\left|f\right|^2}{B}\right)\left(d-\frac{\left|g\right|^2}{A}\right)=|e|^2,\label{Eq.Main.unknown.A.B.a}\\\text{and }\quad
	A\in\left(\frac{\left|g\right|^2}{d},\,a\right),&\: B\in \left(\frac{\left|f\right|^2}{c},\,b\right).\label{Eq.Main.unknown.A.B.b}
	\end{align}
\end{subequations}
Since this is an equation involving only positive variables, we remove the absolute signs for brevity, thereby $\beta_i$ should be understood as $|\beta_i|$ henceforth. (Indeed the condition $M\geq 0$ remains valid if we assume all of $a,b,c,d,e,f,g$ to be positive \cite{Expalin.PSD.of.M}). 

We will first parametrize $A,B$, using the first equation of Eq.~\eqref{Eq.Main.unknown.A.B.a}.  To this end,
\begin{subequations}\label{Exp.A.B.Parameter}
	\begin{align}
	\frac{A(a-A)}{Ad-g^2}&=\frac{Bc-f^2}{B(b-B)}:=k>0.\label{Exp.A.B.Parameter.a}\\
	\Rightarrow\quad A&=\frac{(a-kd)+\sqrt{(a-kd)^2+4kg^2}}{2},\label{Exp.A.B.Parameter.b}\\
	\quad B&=\frac{(kb-c)+\sqrt{(kb-c)^2+4kf^2}}{2k}.\label{Exp.A.B.Parameter.c}
	\end{align}
\end{subequations}
Note that 
\begin{subequations}\label{Exp.a-A.b-B}
	\begin{align}
	a-A&=\frac{(a+kd)-\sqrt{(a+kd)^2-4k(ad-g^2)}}{2},\label{Exp.a-A.b-B.a}\\
	b-B&=\frac{(c+kb)-\sqrt{(c+kb)^2-4k(bc-f^2)}}{2k},\label{Exp.a-A.b-B.b}
	\end{align}
\end{subequations}
where $ad-g^2>0$ and $bc-f^2>0$, thereby $0<A<a$ and $0<B<b$ irrespective of the value of $k>0$. This, together with Eq.~\eqref{Exp.A.B.Parameter.a} shows that Eq.~\eqref{Eq.Main.unknown.A.B.b} is satisfied for any $k>0$.

The final task is to show that there is a positive solution for the equation $(a-A)(b-B)=e^2$ in $k$.
Substituting the values from Eq.~\eqref{Exp.a-A.b-B}, this gives
\begin{align*}
&\left[(a+kd)-\sqrt{(a+kd)^2-4k(ad-g^2)}\right]\left[(c+kb)-\sqrt{(c+kb)^2-4k(bc-f^2)}\right]=4k e^2\\ \Rightarrow\,& (ad-g^2)\left[(c+kb)-\sqrt{(c+kb)^2-4k(bc-f^2)}\right]=e^2\left[(a+kd)+\sqrt{(a+kd)^2-4k(ad-g^2)}\right]\\ \Rightarrow\,& (ad-g^2)\sqrt{(c+kb)^2-4k(bc-f^2)}+e^2\sqrt{(a+kd)^2-4k(ad-g^2)}=(ad-g^2)(c+kb)-e^2(a+kd)
\end{align*}
Squaring both sides twice with appropriate rearrangements and removing the irrelevant factor $16 k(ad-g^2)^2$, the above equation reduces to the following quadratic equation

\begin{subequations}\label{Eq:Quadratic.in.k}
	\begin{align}
	&\qquad\qquad\alpha\, k^2-\beta\, k + \gamma =0,\label{Eq:Quadratic.in.k.a}\\
\text{ where }\qquad	\alpha &= e^2d_2d_4,\\
	\beta&=\Delta \left[\Delta+e^2 \left(a b+c d+4 fg\right)\right]
 + 2 e^4 \left(a b f g+a d f^2+b c g^2+c d f g\right),\\
	\gamma&=e^2d_1d_3,\\
	d_1&:=a(bc-f^2)- ce^2,\,d_2:=b(ad-g^2)- de^2,\\ 
	d_3&:= a(cd-e^2)- cg^2,\, d_4:=b(cd-e^2)- df^2,\\
	\Delta&:=(ad-g^2)(bc-f^2)-e^2(ab+cd-e^2+2fg).
	\end{align} 
\end{subequations}
The quantities $d_1,d_2,d_3,d_4$ are the third-ordered principal minors of $M$ and can be shown strictly positive by either directly evaluating those minors of $M$ from Eq.~\eqref{Exp.Choi.M}, or substituting the values of $a,b,c,d,e,f,g$ in term of $\alpha_i,\beta_i$, the latter showing the strict positivity of $\Delta$, 
\[\Delta = \alpha _1^2 \alpha _4^2 \beta _2^2 \left(\beta _1^2+\beta _3^2\right)+ \alpha _3^2 \alpha _4^2 \beta _1^2 \left(\beta _2^2+\beta _1^2 r^2\right)+r^2 \left(\alpha _2 \alpha _3 \beta _1^2-\alpha _1^2 \beta _2 \beta _3\right)^2>0.\]
Therefore,  $\alpha>0,\,\beta>0,\,\gamma>0$. Noticing that the discriminant \cite{Explain.D>0}\[ D:=\beta^2-4\alpha\gamma=\Delta \left(\Delta+4e^2f g\right)\left[\left(ad-g^2\right)\left(bc-f^2\right)-e^4\right]^2>0,\]
it follows that the two roots of the quadratic Eq.~\eqref{Eq:Quadratic.in.k.a} are positive (and distinct).\hfill $\square$

%============================== Section 3 ==========================
\section{IO vs. SIO}
A general (without any restriction like incoherentness) single qubit channel can optimally be decomposed into four Kraus operators. So, incoherentness makes no difference for qubit channels in terms of optimal Kraus decomposition. Note, however, that it no way means that a general channel could be decomposed into incoherent Kraus operators. A 2-by-2 unitary with non-vanishing entries (even in any single column) is a simple example of a channel which is not incoherent. This is also true for any finite dimension. A $d$-dimensional incoherent unitary is a general permutation (that is, the entries are arbitrary phases, $e^{i\phi}$, not necessarily 1). Thus, all IO unitaries (channels with Kraus rank one) are necessarily SIO. For non-unitary channels, this is not true already in qubit level.
\begin{proposition}\label{Prop:IO.not.SIO}
There are (qubit-) incoherent channels which are not strictly incoherent.
\end{proposition}
\noindent\textbf{Proof:} As an example, consider the following incoherent channel
\begin{equation}
\label{Eq:IO.not.SIO} \Lambda=\left\{\frac{1}{2}\begin{pmatrix}
1&1\\0&0
\end{pmatrix},\,\frac{1}{2}\begin{pmatrix}
0&0\\1&-1
\end{pmatrix},\,\frac{1}{2}\begin{pmatrix}
1&0\\0&1
\end{pmatrix},\,\frac{1}{2}\begin{pmatrix}
0&1\\1&0
\end{pmatrix}\right\}.
\end{equation}
If it were SIO, there must be a unitary converting those four Kraus operators via Eq.~\eqref{Eq:Equivalence.of.Kraus.operators} to the standard SIO form given in Eq.~\eqref{Pres:canonical.SIO}. Comparing the last two operators, the fourth and third (unnormalized) row of the unitary must be $(1,-1,-1,-1)x$ and $(1,1,1,-1)y$ for some nonzero $x,y\in\CC$. The orthogonality of those two row vectors with the first row implies that the first row must be $(a,b,-b,a)$ with $a,b\in\CC$. However, the resulting first Kraus operator is then \[\begin{pmatrix}
\frac{a-b}{2}&a\\\frac{a+b}{2}&-b
\end{pmatrix}.\] In order it to be an SIO, we must have $a=0=b$, an impossibility as a unitary cannot have a zero row. \hfill$\square$ 

The four Kraus operators in Eq.~\eqref{Eq:IO.not.SIO} are linearly independent. Hence the channel cannot be reduced further to have three Kraus operators. 

In the qubit case, it is easy to characterize all IO with two Kraus operators which are also SIO.
\begin{proposition}\label{Pro:rank.2.IO.not.SIO}
All qubit IO with two Kraus operators are essentially SIO, except the following canonical one \begin{equation}\label{Eq:rank.2.IO.not.SIO}
\left\{\begin{pmatrix}
\cos\theta&\sin\theta e^{i\phi}\\0&0
\end{pmatrix},\,\begin{pmatrix}
0&0\\\sin\theta& -\cos\theta e^{i\phi}
\end{pmatrix}\right\},\quad 0<\theta<\frac{\pi}{2},\,\phi\in\RR.
\end{equation}
\end{proposition}
\noindent\textbf{Proof:} There are only three IOs to verify: \[\left\{\begin{pmatrix}
*&*\\0&0
\end{pmatrix},\,\begin{pmatrix}
*&*\\0&0
\end{pmatrix}\right\},\;\left\{\begin{pmatrix}
0&0\\ *&*
\end{pmatrix},\begin{pmatrix}
0&0\\ *&*
\end{pmatrix}\right\},\;\left\{\begin{pmatrix}
*&*\\0&0
\end{pmatrix},\,\begin{pmatrix}
0&0\\ *&*
\end{pmatrix}\right\},\] 
where * denotes an arbitrary non zero complex number (subject to the restriction of forming a channel).
The first one could be parametrized as \[\left\{\begin{pmatrix}
\cos\theta&\sin\theta e^{i\phi}\\0&0
\end{pmatrix},\,\begin{pmatrix}
\sin\theta& -\cos\theta e^{i\phi}\\0&0
\end{pmatrix}\right\},\quad 0<\theta<\frac{\pi}{2},\,\phi\in\RR,\]
and hence the unitary \[U=\begin{pmatrix}
\cos\theta&\sin\theta\\\sin\theta&-\cos\theta
\end{pmatrix}\] transforms it to an SIO. Similarly, the second one is also an SIO. For the last one, however, one verifies that no $4$-by-$4$ unitary could transform it to an SIO given in Eq.~\eqref{Pres:canonical.SIO}.\hfill$\square$ 
 
Note that the channel $\{K_1,K_2\}$ in Eq.~\eqref{Eq:rank.2.IO.not.SIO} trivially extends to higher dimension: the IO $\{K_1\oplus0,\,K_2\oplus0,\,|2\ran\lan2|,\dotsc,|d\ran\lan d|\}$ is not an SIO.

%============================== Section 4 ==========================
\section{Kraus rank vs. IO rank vs. SIO rank} In analogy with Kraus rank, we say the minimum number of (S)IO Kraus operators of an (S)IO channel its (S)IO rank. Since SIO $\subseteq$ IO, it follows that for an SIO,
\[\text{ Kraus rank }\leq \text{ IO rank }\leq \text{SIO rank },\]
where the inequalities are expected to be strict for higher dimension. In qubit case, however, the two terminal ranks are same. 

\begin{proposition}\label{Pro:SIO.rank=Kraus.rank.for.qubit}
	For all qubit SIO, the SIO rank equals to Kraus rank. This is not necessarily true in higher dimension.
\end{proposition} 
\noindent\textbf{Proof:} The CJS matrix for a qubit SIO, which without loss of generality given by Eq.~\eqref{Pres:canonical.SIO}, can be decomposed as
\begin{equation} 2M=\begin{pmatrix}
a_3^2 &.&.&.\\.&.&.&.\\.&.&a_4^2&.\\.&.&.&.
\end{pmatrix}+\begin{pmatrix}
a_1^2&.&.&a_1b_1^*\\. & . & . &.\\ .& .& .&.\\ a_1b_1&.&.&|b_1|^2
\end{pmatrix}+\begin{pmatrix}
. & .&. &.\\.&|b_2|^2&a_2b_2&.\\.&a_2b_2^*&a_2^2&. \\.&.&. &.
\end{pmatrix}.
\end{equation} 
This shows that the Kraus rank of the SIO is four, unless $b_1b_2=0$ or some of the Kraus operators vanish  entirely. By the channel (normalization) condition, $b_1,b_2$ cannot vanish both. If $b_1=0$, the two operators $\binom{a_1~0}{0~~0}$ and $\binom{a_3~0}{0~~0}$ are scalar multiple of each other and so one of them could be made zero, thereby diminishing the SIO rank to match with Kraus rank. Similar reasoning applies to the case $b_2=0$. If exactly $k$ number of Kraus operators vanish, both the Kraus rank and SIO rank  also diminish by $k$. Thus the two ranks are always same. 

 To show that this is not necessarily true in higher dimensions, consider the (unnormalized) qutrit SIO consisting of the six $3$-by-$3$ permutation matrices as Kraus operators. The CJS matrix has rank five thereby Kraus rank is $5$. However, there is no $6$-by-$6$ unitary $U$ which could reduce the SIO rank from $6$. To prove this latter claim, we first note that $a=(1, -1, -1, 1, 1, -1)$ must be an unnormalized row of $U$ (so that one of the transformed Kraus operators vanishes). Then we verify that among the resulting Kraus operators, no two can have the same form of any of the six-permutations; for example, if two have the same form as the identity permutation, then the two corresponding rows of $U$ must be $b=(x_1,x_2,x_2,-x_2,-x_2,x_2)$ and $c=(y_1,y_2,y_2,-y_2,-y_2,y_2)$. However, $a,b,c$ has to be mutually orthogonal, forcing one of $b,c$ to be zero, an impossibility. All the other cases of repeating one form can be discarded by similar arguments. So, the transformed non-vanishing five Kraus operators must be of the form of six permutations, in particular two of them must have the form of $\{1,2\}$, or $\{2,3\}$, or $\{3,1\}$. But, like the previous cases, all these lead to a zero row and hence it is impossible to reduce the SIO rank.   \hfill$\square$    

For IO channel, the two ranks can differ already at qubit level.
\begin{proposition}\label{Prop:IOrank.neq.Krausrank}
	There are (qubit-) incoherent channels with Kraus rank $<$ IO rank.
\end{proposition}
\noindent\textbf{Proof:} As an example, consider the following incoherent channel
\begin{equation}
\label{Eq:IOrank.neq.Krausrank} \Lambda=\left\{\frac{1}{2}\begin{pmatrix}
1&1\\0&0
\end{pmatrix},\,\frac{1}{2}\begin{pmatrix}
0&0\\1&-1
\end{pmatrix},\,\frac{1}{2}\begin{pmatrix}
1&0\\0&-1
\end{pmatrix},\,\frac{1}{2}\begin{pmatrix}
0&1\\1&0
\end{pmatrix}\right\}.
\end{equation}
This channel has Kruas rank $3$, but IO rank $4$.

%============================== Section 5 ==========================
\section{Discussion and conclusion} Our primary aim in this work was to find a minimal description of qubit incoherent operations. We have achieved this, as detailed in the main result section, by showing that any qubit IO could be decomposed into four incoherent Kraus operators. This shows that at most eight real parameters are needed to simulate any qubit IO.

\begin{figure}[t]
	\centering
	\includegraphics[width=0.8\columnwidth]{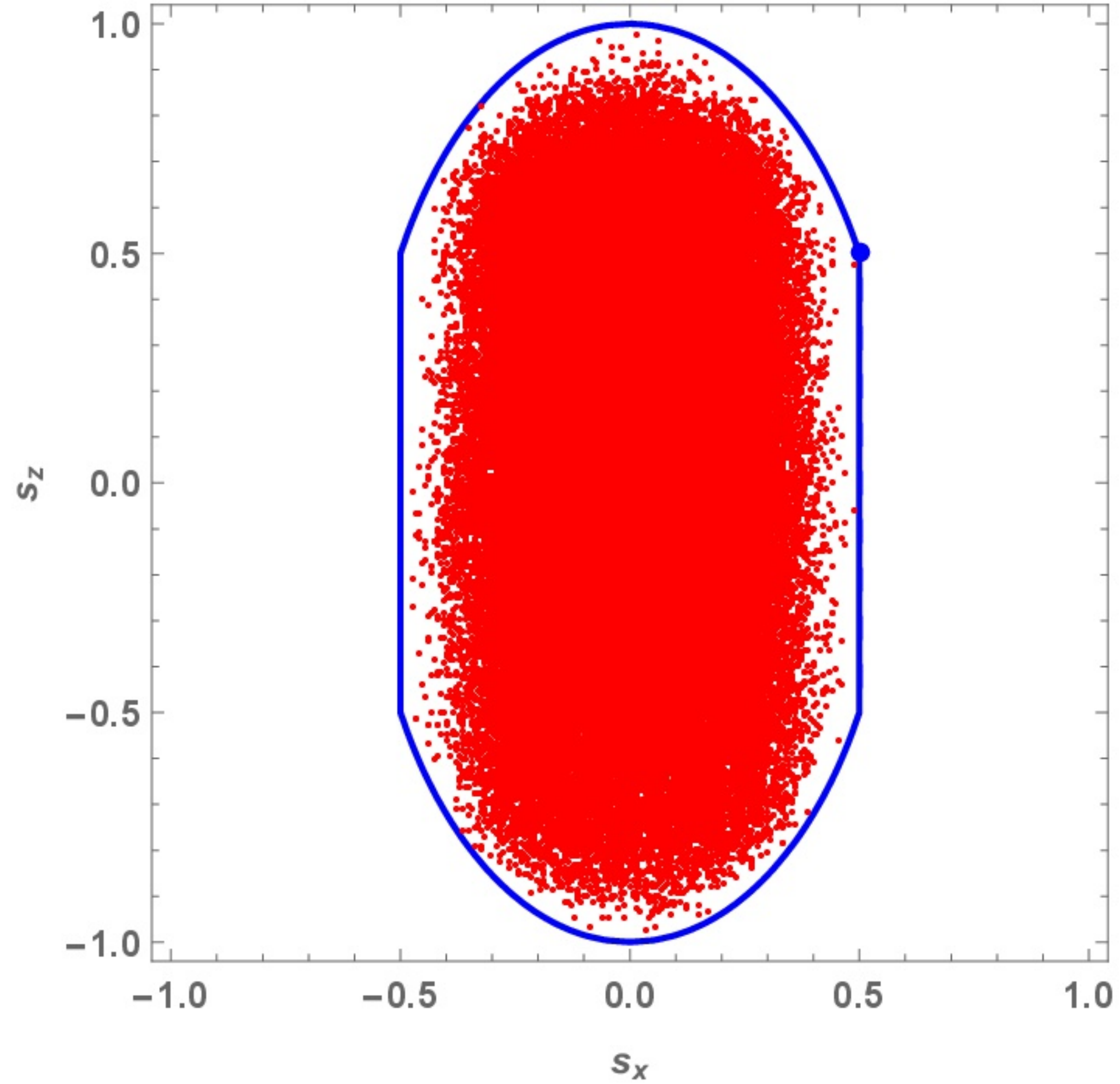}
	\caption{\label{fig:achieaval.IO} (color online) Numerical simulation of achievable region for single-qubit IO. The red colored area (each dot represents a state) shows the projection of the achievable region in the $x$-$z$ plane for initial Bloch vector $(0.5,0,0.5)^{T}$ [blue dot], while the blue thick curve is the exact analytical boundary of the achievable region. In this figure we have simulated $10^5$ random IO channels, given by the canonical form in Eq.~\eqref{Pre.Canonical.IO.Optimal}, in Mathematica---\texttt{$r$ = RandomReal[$\{1, 2\}$],  $\alpha=$Normalize[RandomReal[$\{~\}$, 3]], $\beta=$Normalize[RandomComplex[$\{-2-2 i,\,2+2 i\}$,\,3]]}, and replacing $\alpha_1,\beta_1$ by $\alpha[[1]]/\sqrt{1+r^2}$, $\beta[[1]]/\sqrt{1+r^2}$ respectively. }
\end{figure}

As an application of this optimal decomposition, we have numerically simulated the set of reachable states, from a given initial qubit state by all possible qubit IOs. The results are good approximations of the exact analytic achievable regions from \cite{Streltsov+3.PRL.2017}. We have depicted one such simulation in Fig.~\ref{fig:achieaval.IO}. Evidently, the figure is quite suggestive as many points reach the exact analytic boundary.     

 Regarding state conversion under various models of free operations (see e.g. \cite[Table~II]{Streltsov+2.RMP.2017}) in coherence theory, it is known that all hierarchies collapse for qubit state conversion \cite{Streltsov+3.PRL.2017,Chitambar+1.PRL.2016}. The achievable regions remains same even for probabilistic IO and SIO \cite{Theurer+2.Ar.2018}. However, the optimal decomposition serves much higher purpose, as it allows to apply IO only one part of a multipartite systems, which is the main theme in distributed scenario \cite{Streltsov+3.PRX.2017}.
 
To conclude, we have shown that every IO can be decomposed into at most four incoherent Kraus operators and there are some requiring exactly four. The problem of finding minimum number of (strictly) incoherent Kraus operators for (S)IO beyond qubit systems remains open.

\medskip

\noindent\textbf{Acknowledgements.} We thank Alexander Streltsov and Preeti Parashar for helpful discussions. We acknowledge financial support from ERC grant OSYRIS (ERC-2013-AdG Grant No.\ 339106), EU grant QUIC (H2020-FETPROACT-2014 Grant No. 641122), the European Social Fund, the Spanish MINECO grant FISICATEAMO (FIS2016-79508-P), the Severo Ochoa Programme (SEV-2015-0522), MINECO CLUSTER (ICFO15-EE-3785), the Generalitat de Catalunya (2014 SGR 874 and CERCA/Program), the
Fundaci\'{o} Privada Cellex, and the National Science Centre, Poland-Symfonia (Grant No. 2016/20/W/ST4/00314).


\begin{thebibliography}{30}
\bibitem{Expalin.PSD.of.M} Notice that for the Hermitian matrix \[X=\begin{pmatrix}
a & e & 0 & g \\
e^* & b & f & 0 \\
0 & f^* & c & -e \\
h^* & 0 & -e^* & d 
\end{pmatrix},\]
$X\geq 0$ implies $\tilde{X}\geq 0$, where $\tilde{X}$ is the matrix obtained from $X$ by keeping the same diagonals and replacing the off-diagonal elements $e,f,-e,g$ by $|e|,|f|,-|e|,|g|$ respectively . This is due to the fact that in the condition for PSDness of $X$, phases appear only in the determinant condition, which reads 
\begin{align*}
0 &\leq \det\left(X\right)\\
&=\left(ad-|g|^2\right)\left(bc-|f|^2\right)-|e|^2\left(ab+cd-|e|^2\right)+2\Re\left[e^2fg^*\right]\\
&\leq \left(ad-|g|^2\right)\left(bc-|f|^2\right)-|e|^2\left(ab+cd-|e|^2\right)+2|e|^2|f||g|\\
&=\det\left(\tilde{X}\right). 
\end{align*} 

The caveat is that $X$ and $\tilde{X}$ could be entirely different matrix (e.g., with different rank). So, in our problem, we could not use the decomposition of $2\tilde{M}$ to claim the same about $2M$.
\bibitem{Explain.D>0} The inequality $\left(ad-g^2\right)\left(bc-f^2\right)-e^4>0$ could be shown either by substituting $a,b,c,$ etc. in terms of $\alpha_i,\beta_i$, or by writing the left hand side as $\Delta+e^2[(ab-e^2)+(cd-e^2)+2fg]$.
\end{thebibliography}
\end{document}